\documentclass[showpacs,preprintnumbers,twocolumn,amsmath,amssymb,floatfix,,prd]{revtex4}

\usepackage{graphicx}
\usepackage{dcolumn}
\usepackage{bm}
\usepackage{hyperref}
\hypersetup{colorlinks,citecolor=blue}
\usepackage{color}

\begin{document}

\title{Compact stars with exotic matter}

\author{Ksh. Newton Singh}
\email{ntnphy@gmail.com}
\affiliation{Department  of  Physics,  National  Defence  Academy, Khadakwasla,  Pune-411023,  India\\
Department of Mathematics, Jadavpur  University,  Kolkata 700032,  India}

\author{Amna Ali}
\email{amnaalig@gmail.com}
\affiliation{Department of Mathematics, Jadavpur University, Kolkata 700032, West Bengal, India}

\author{Farook Rahaman}
\email{rahaman@associates.iucaa.in,}
\affiliation{Department of Mathematics, Jadavpur University, Kolkata 700032, West Bengal, India}

\author{Salah Nasri}
\email{snasri@uaeu.ac.ae}
\affiliation{Physics Department, UAE University, POB 17551, Al Ain, United Arab Emirates \\
International Center for Theoretical Physics, Trieste, Italy.}

\date{\today}

\begin{abstract}
In this paper we employ Tolman VII solution  with exotic matter that may be present in the extremely dense core of compact objects. The Tolman VII solution, an exact analytic solution to the spherically symmetric, static Einstein field equations with a perfect fluid source, has many characteristics that make it interesting for modeling high density stellar astronomical
objects. For our purpose we use generalized non-linear equation of state which may incorporate exotic matter along with dust, radiation and dark energy. The equation of state (EoS) has a linear contribution comprising quark/radiation matters and a nonlinear contribution comprising  dark energy/exotic matters. The amount of exotic matter contain can be modify by a parameter $n$  which can be linked to adiabatic index. As $n$ increases the exotic contribution increases  which stiffens the EoS. The physical properties of the model such as pressure, density, mass function, surface redshift, gravitational redshift are examined and the stability of the stellar configuration is discussed in details. The model has promising features as it satisfies all energy conditions and is free from central singularities. The  $M-R$ relation is constructed analytically and the maximum mass and its corresponding radius is determined using the exact solutions and is shown to satisfy various observational stellar compact stars.

\end{abstract}
\keywords{ Exact solutions  \and  Relativistic stars: structure \and Stability,}
\pacs{04.40.Nr, 04.20.Jb, 04.20.Dw, 04.40.Dg}

\maketitle

\section{Introduction}\label{Sec1}
The advances in astrophysical observations has lead to great interest in the study of composition of the astrophysical compact object. Moreover these objects contains compressed ultra-dense nuclear matter in their interiors which make them  superb astrophysical laboratories for a wide range of intriguing physical studies. Traditionally the term compact objects or compact stars refers collectively to white dwarfs, neutron stars, and black holes.
Compact stars are also called the stellar remnants as they are often the endpoints of  catastrophic astrophysical events such as supernova explosions and binary stellar collisions. The state and type of a stellar remnant depends primarily on the composition and properties of the dense matter of the star.  However, due to lack of knowledge of the extreme condition and its complex composition it is a difficult task to determine the
exact equation of state(EoS) to represent compact stars. Various astrophysical observations measure masses and radii of compact stars \cite{Pons, Drake,Walter,Cottam}, which in turn, constrain the EoS of dense matter in compact stars. For example the observation of 2-solar mass neutron stars \cite{Demorest, Antoniadis} suggests that the EoS for compact stars needs to be sufficiently stiffer than the normal nuclear matter to
accommodate the large mass. This enables one to think of a stable mass configurations with “exotic” matter in
their interiors. In case of low mass compact stars too, the core matter density is much larger than the normal matter. Due to
extreme density, nuclear matter may consist not only of
nucleons and leptons but also several exotic components in
their different forms and phases such as Bose-Einstein condensates of strange mesons\cite{Kaplan,Nelson,Schaffner,Glendenning, Banik}, hyperons,
and baryon resonances \cite{Glendenning2}, as well as strange quark matter \cite{Prakash, Farhi,Schertler}. 

Constructing the EoS of matter above the nuclear saturation density, relevant for the description of compact stars, is a vast arena for research.
For a  proposed EoS, the study of physical features of relativistic spheres like compact objects
in general relativity is done by finding the
exact analytic solutions for static Einstein field equations and imposing conditions for physical acceptability. However, it is a daunting task to obtain explicit analytical solutions of Einstein's field equation on account of their complicated and non-linear nature. Karl Schwarzschild obtained the first exact solution of Einstein's field equations \cite{Sch}. The number of
valid, exact solutions has been growing since then
and are extensively used in the studies of neutron stars and black hole formation \cite{Ray, Felice}. 
 Exact solutions for modelling more realistic relativistic fluids include Buchdahl \cite{Buchdahl} and Tolman VII \cite{Tolman, Hobill} solutions. In pirticular, Tolman VII solution is stable for a large range of compactness (ratio of mass over the radius)\cite{Negi1,Negi2} and is used to study various problems related to very compact object \cite{Neary}.
 For better understanding of the compact objects these analytical solutions with various EoS are considered in the literature.  In particular the linear EoS is considered to model charged or neutral  anisotropic relativistic fluid with strange quark matter \cite{Mak, Sharma, Esculpi, Komathiraj, Takisa1, Takisa2, Rahaman1, Kalam, Maharaj}. 
 
Usually EoS of dense matter including exotic phases  are constructed using relativistic field
theoretical models and chiral models. It was noted that
the appearance of exotic forms of matter in the high density regime resulted in kinks in the
EoS \cite{Schaffner}, which resulted discontinuity in the speed of sound. This has a great implication on determining the stability of the star. Using fundamental particle physics, quark matter at high density is studied which  leads to the MIT-bag model for strange stars \cite{Chodos}. This model added a correction term to the usual classical barotropic EoS called the Bag constant and assumes that the free quarks in stellar configuration is confined in a bag characterized by a vacuum energy density equal to the bag constant. As it is well established from cosmological observations that around $70\% $ of the energy budget of universe is invisible dark energy, there is a growing interest to understand whether a stable configuration of compact can be made up of dark energy. For this various dark energy EoS has been employed to model compact objects ranging from quark stars through to neutron stars \cite{Rahaman2}.
 
The phenomenon of late time cosmic acceleration \cite{Perlmutter,Riess} can be understood by incorporating dark energy as an exotic relativistic fluid with large negative pressure fills the whole universe.  There is also an  alternative view  according to which current cosmic acceleration is an artifact of modification of gravity at large scale rather than the consequence of dark energy. Tons of  theoretical approaches have been employed to explain the evolution of universe in the light of cosmological observations. More recently the theory of ever exiting or emergent universe \cite{Ellis} was formulated under which it may be possible to build models which avoid the quantum regime for space-time, nevertheless share the novel features of the standard big-bang model. This scenario can be treated as alternative inflationary model  within the standard big bang framework which incorporates an asymptotically Einstein static universe in the past and evolves to an accelerating universe subsequently. The emergent universe scenario and has been studied recently in different theories of modified gravity\cite{Cai,Parisi,Zhang}, Brane world models \cite{Banerjee}, Brans-Dicke \cite{Paul}. In the framework of $f(R)$ gravity, Mukherjee et al. \cite{Mukherjee} pointed out that the Starobinsky model, the original as well as the modified version, permits solutions portraying an emergent universe. Subsequently, a general framework for such a scenario in general relativity was proposed \cite{Dadhich} with
a different constituents of matter that are prescribed by an non linear EoS : $p=A\rho-B \rho^{1/n}$, where $n,~ A$ and $B$ are constants. For $B > 0$ and $n=2$ the possible primordial compositions of universe has been suggested that are permitted by the EoS. It admits existence of exotic matter such as cosmic strings, domain wall, quark matter and dark energy in addition to radiation and dust.

The purpose of the paper is to model the  stellar Compact star, with aforesaid EoS and determine the physical stability by studying its exact solutions. The article is organized as follows. In Sec II we consider a spherical symmetric metric and present the Einstein's equations for anisotropic fluid distributions. In Sec. III. We employ the Tolman VII equation with above EoS. and obtain the expressions for density and pressures in Sec. III. Next, in Sec VI we investigate non-Singular nature of the solutions. In Sec.V, from the Boundary conditions we determine the constant parameters of the model. In Sec. VII we discuss the stability of the model and obtain the analytic results. Finally, Sec. VIII is devoted to concluding remarks of the study.

\section{Interior space-time and field equations}\label{2}

The interior space-time line element for an uncharged, static and spherically symmetric fluid is given by:

\begin{equation}
ds^{2}=e^{\nu(r)}dt^{2}-e^{\lambda(r)}dr^{2}-r^{2}\left(d\theta^{2}+\sin^{2}\theta d\phi^{2} \right) \label{met}
\end{equation}
where $\nu$ and $\lambda$ are functions of the radial coordinate `$r$' only.\\
The interior of a star is often modelled as a perfect fluid, which requires the pressure to be isotropic. However, theoretical studies indicate that, at extremely high densities, deviations from local isotropy may play an important role \cite{Krsna}.  It was argued \cite{rude} that at high density regimes ($\rho>10^{15} ~g/cm^3$) the nuclear matter interacts relativistically as a result of which nuclear matter may have anisotropic features. The numerical calculations of Barreton \& Rojas \cite{barr} suggests that the exotic phase transitions that occurs during the process of gravitational collapse \cite{coll,itoh} such as  pion condensed state \cite{hart}, anisotropic stress tensor associated type II superconductor \cite{jone,eass}, solid core \cite{rude}, type P superfluid \cite{ruff} etc. may also induce anisotropy. Therefore, assuming an anisotropic fluid distribution the Einstein's field equations can be written as
\begin{eqnarray}
R^\mu_\nu-{1\over 2}g^\mu_\nu R &=& -{8\pi} \big[(p_t +\rho c^2)v^\mu v_\nu-p_t g^\mu_\nu+(p_r-p_t) \nonumber \\
&& \chi_\nu \chi^\mu \big] \label{fil}
\end{eqnarray}
where the symbols have their usual meanings.\\
For the space-time \eqref{met}, the field equations reduces to
\begin{eqnarray}
\frac{1-e^{-\lambda}}{r^{2}}+\frac{e^{-\lambda}\lambda'}{r} &=& 8\pi\rho \label{dens}\\
\frac{e^{-\lambda}-1}{r^{2}}+\frac{e^{-\lambda}\nu'}{r} &=& 8\pi p_{r} \label{prs}\\
e^{-\lambda}\left(\frac{\nu''}{2}+\frac{\nu'^{2}}{4}-\frac{\nu'\lambda'}{4}+\frac{\nu'-\lambda'}{2r} \right) &=& 8\pi p_t\, \label{prt}
\end{eqnarray}
where $(')$ denotes derivative with respect to radial coordinate r.
We define that the measure of anisotropy: $\Delta = p_t - p_r$.\\

To study the static spherically symmetric configurations with
anisotropic matter distribution we adopt the following variable transformations:
\begin{equation}\label{tr}
x=r^{2},~~~~~~~Z(x)=e^{-\lambda(r)}~~~~~ \text{and}~~~~ y^{2}(x)=e^{\nu(r)},
\end{equation}
the field equations (\ref{dens})-(\ref{prt})then takes the form
\begin{eqnarray}
8\pi\rho &=& \frac{1-Z}{x}-2\dot{Z} \label{rho1}\\
8\pi p_r &=& 4Z\frac{\dot{y}}{y}+\frac{Z-1}{x} \label{pr1}\\
8\pi\Delta &=& 4xZ\frac{\ddot{y}}{y}+\dot{Z}\left(1+2x\frac{\dot{y}}{y}\right)+\frac{1-Z}{x} \label{delta1}\\
p_t &=& p_r+\Delta \label{pt1}
\end{eqnarray}
where $\dot{Z}=dZ/dx$ and $\ddot{Z}=d^2Z/dx^2$.\\

To proceed further, we assume $Z(x)$ as Tolman VII and a nonlinear equation of state as
\begin{eqnarray}
Z(x) = 1-ax+bx^2~~~\mbox{and} ~~~p_r=A\rho-B \rho^{1/n}. \label{zeo}
\end{eqnarray}
Here $a,~b,~A$, $B$ and $n$ are arbitrary constants. \\

 These constants are not restricted to specific values. For $A=0,~\gamma = 1/n,~B=-K$ gets $p = K \rho^\gamma$ (polytropic EoS); $A=1/3,~B=4\mathcal{B}/3,~$ gives $p=(\rho-4\mathcal{B})/3$ (the MIT model); $B=0,~A=1$ gets $p=\rho$ ( Zeldovich's stiff fluid); $A=1/3,~B=0$ yields radiation ($p=\rho/3$) etc. Hence, for $0<A \le 1$, the first term of the EoS represents normal/stiff matter/relativistic matter depending on the values of $A$. The second term is the non-linear extension and by choosing $B>0$, we ensure that it is the source of dark energy and exotic matter. Since we have chosen $n>1$, the quantity $p_r/\rho$ given below
\begin{equation}
{p_r \over \rho } = A-{B \over \rho^{1-1/n}}
\end{equation}
tends to $A$ as $\rho \rightarrow \infty$. This means that the contribution of dark energy component is minimum at the center.
Therefore, the dense core is populated with exotic and stiff fluid. The density at which the second term in the above equation vanishes depends on parameter $n$ thereby determining the amount of exotic matter in the stellar configuration. For large $n$ the exotic component is more than for low $n$. As exotic contribution  stiffens the EoS therefore, the parameter $n$ is directly linked with the stiffness of EoS. For $n=2$, $B>0$ and $-1 \leq A \leq 1$, it is shown that the EoS satisfies various constituents of  universe \cite{Dadhich}, which includes exotic matter, cosmic strings, dark energy, dust, radiation and stiff matter. We intend to analyze the EoS for a variable $n$ and deduce the other constants using the boundary condition to fit for observed compact stars i.e. the X-ray pulsar LMC X-4.

\begin{figure}[t]
\centering
\includegraphics[width=7cm,height=4.5cm]{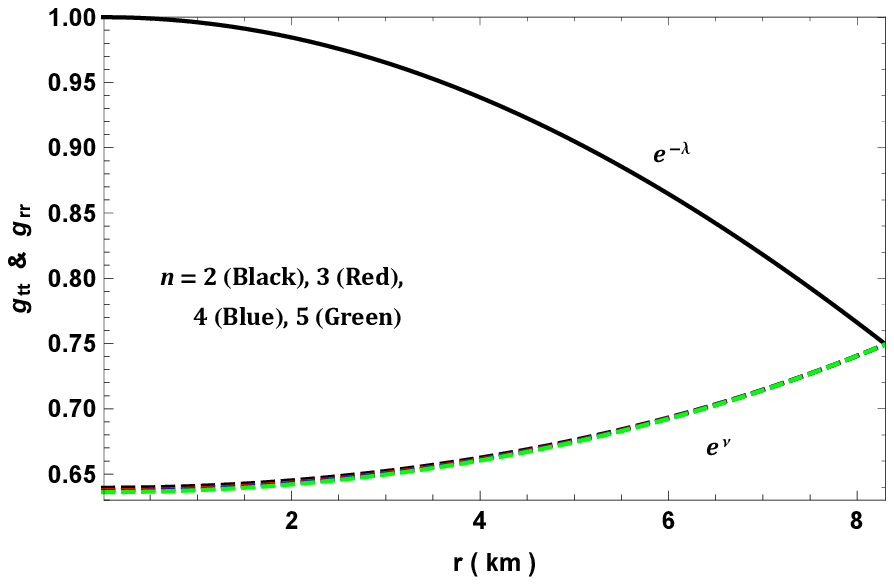}
\caption{Variation of metric functions with $r$ for LMC X-4 assuming $M=1.04M_\odot,~R=8.301~km, ~a = 0.0039$ and $A = 0.7$.}\label{fi1}
\includegraphics[width=7cm,height=4.5cm]{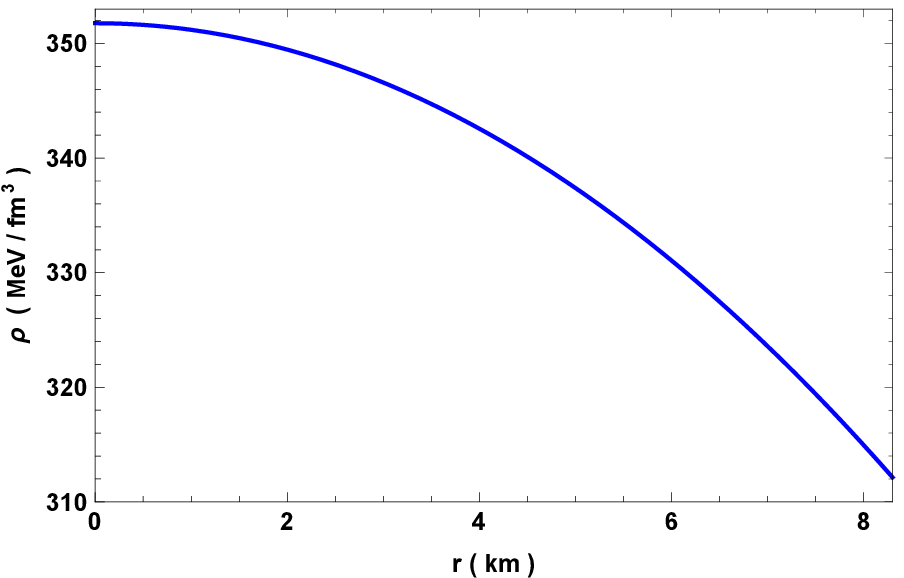}
\caption{Variation of energy density with $r$ for LMC X-4 assuming $M=1.04M_\odot,~R=8.301~km, ~a = 0.0039$ and $A = 0.7$.}\label{fi2}
\end{figure}

\begin{figure}[t]
\centering
\includegraphics[width=7cm,height=4.5cm]{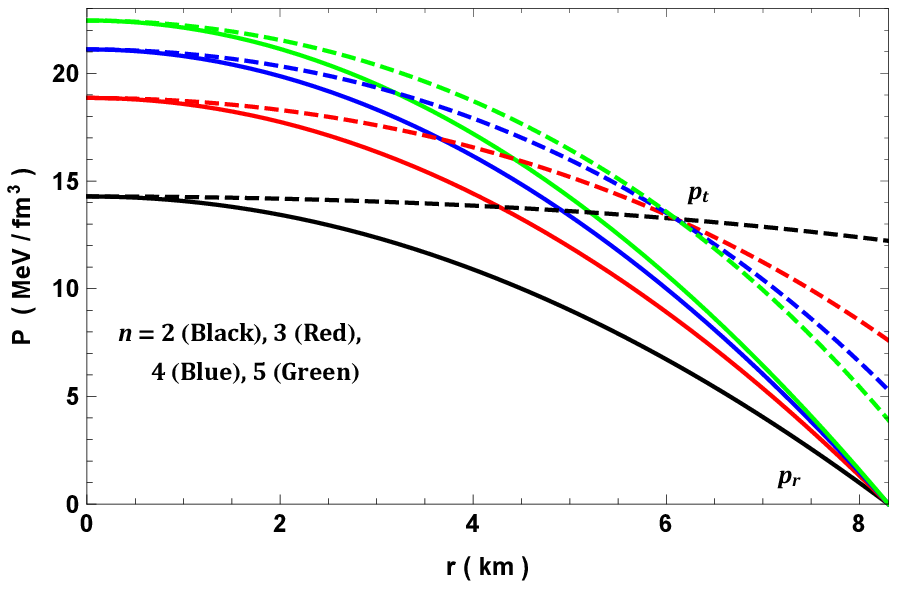}
\caption{Variation of pressure with $r$ for LMC X-4 assuming $M=1.04M_\odot,~R=8.301~km, ~a = 0.0039$ and $A = 0.7$.}\label{fi3}
\includegraphics[width=7cm,height=4.5cm]{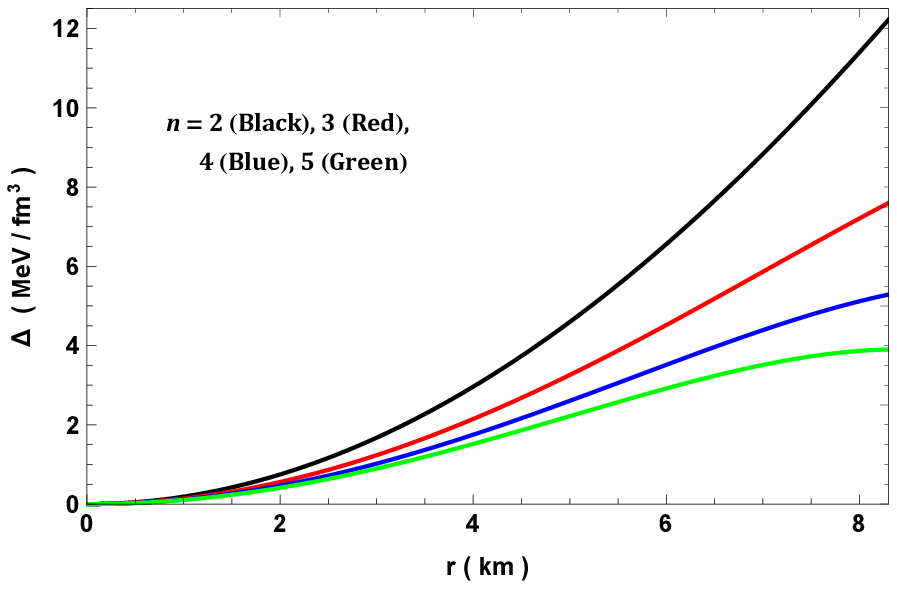}
\caption{Variation of anisotropy with $r$ for LMC X-4 assuming $M=1.04M_\odot,~R=8.301~km, ~a = 0.0039$ and $A = 0.7$.}\label{fi4}
\end{figure}

\section{A generalized solution}\label{3}

Solution of the field equations depend on the metric function $\nu$ and $\lambda$. As discussed above we choose Tolman VII $g_{rr}$ satisfying a nonlinear EoS. Using these in eqn. (\ref{tr}) we get
\begin{eqnarray}
y(x) &=& F \left(1-a x+b x^2\right)^{-\frac{5 A}{8}-\frac{1}{8}} \exp \bigg[\frac{a (A+1)}{4 \sqrt{4 b-a^2}} \nonumber \\
&& \tan ^{-1}\left(\frac{2 b x-a}{\sqrt{4 b-a^2}}\right)+ \frac{2^{\frac{n-4}{n}} 5^{1/n} Bn \pi ^{\frac{n-1}{n}} \tau(x) }{\sqrt{a^2-4 b}} \nonumber \\
&& (3 a-5 b x)^{1/n} \bigg] ~,\label{ela}
\end{eqnarray}

where $F$  is constant of integration and 
\begin{eqnarray}
\tau(x) &=&  \chi(x)  \left(\frac{3 a-5 b x}{a-\sqrt{a^2-4 b}-2 b x}\right)^{-1/n}-\xi(x) \nonumber \\
&&  \left(\frac{3 a-5 b x}{\sqrt{a^2-4 b}+a-2 b x}\right)^{-1/n} \nonumber \\
\chi(x) &=& \, _2F_1\left[-\frac{1}{n},-\frac{1}{n};\frac{n-1}{n};\frac{5 \sqrt{a^2-4 b}-a}{5 \left(\sqrt{a^2-4 b}+a-2 b x\right)}\right] \nonumber 
\end{eqnarray}
\begin{eqnarray}
\xi(x) &=& \, _2F_1\left[-\frac{1}{n},-\frac{1}{n};\frac{n-1}{n};\frac{5 \sqrt{a^2-4 b}+a}{5 \left(\sqrt{a^2-4 b}-a+2 b x\right)}\right].\nonumber
\end{eqnarray}
Using the above equations we deduce the physical properties of the compact star such as density, radial and transverse pressure and anisotropic factor from eqns. (\ref{rho1}), (\ref{pr1}) and (\ref{delta1}) which leads us to,
\begin{eqnarray}
\rho(x) &=& \frac{3 a-5 b x}{8 \pi } \label{den}\\
p_r(x) &=& A\rho-B \rho^{1/n} \label{pre1}\\
\Delta(x) &=& \frac{2^{-\frac{6}{n}-5} \pi ^{-\frac{n+2}{n}} x}{n (3 a-5 b x) \left(a x-b x^2-1\right)} \nonumber \\
&& \bigg[3 a^2 n (8 \pi )^{1/n} f_1(x)-9 a^3 \left(3 A^2+4 A+1\right)  \nonumber \\
&& 64^{1/n} n \pi ^{2/n}-a \left\{f_2(x)+f_3(x)\right\}+5 b \big\{f_4(x) \nonumber \\
&& +f_5(x)+f_6(x)\big\} \bigg] \\
p_t (x) &=& p_r(x)+\Delta(x) ~,
\end{eqnarray}
where,
\begin{eqnarray}
f_1(x) &=& 2 A \left[24 \pi  B (3 a-5 b x)^{1/n}+19 b (8 \pi )^{1/n} x\right]+32 \pi  B \nonumber \\
&&  (3 a-5 b x)^{1/n}+45 A^2 b (8 \pi )^{1/n} x+13 b (8 \pi )^{1/n} x \nonumber\\
f_2(x) &=& 192 \pi ^2 B^2 n (3 a-5 b x)^{2/n}+b B 2^{\frac{3}{n}+4} (19 n-10) \nonumber\\
&& \pi ^{\frac{1}{n}+1} x (3 a-5 b x)^{1/n}+225 A^2 b^2 64^{1/n} n \pi ^{2/n} x^2 \nonumber\\
&& +55 b^2 64^{1/n} n \pi ^{2/n} x^2 \nonumber\\
f_3(x) &=& 15 A b 2^{\frac{3}{n}+2} n \pi ^{1/n} \Big[8 \pi  B x (3 a-5 b x)^{1/n}+2^{\frac{n+3}{n}}b \nonumber\\
&& \pi ^{1/n} x^2-(8 \pi )^{1/n}\Big] \nonumber\\
f_4(x) &=& b B 2^{\frac{3}{n}+4} (3 n-2) \pi ^{\frac{1}{n}+1} x^2 (3 a-5 b x)^{1/n}+25 A^2  \nonumber\\
&&  b^2 64^{1/n} n \pi ^{2/n} x^3+5 b^2 64^{1/n} n \pi ^{2/n} x^3 \nonumber\\
f_5(x) &=& 5 A b 2^{\frac{n+3}{n}} n \pi ^{1/n} x \Big[8 \pi  B x (3 a-5 b x)^{1/n}+b \nonumber\\
&& (8 \pi )^{1/n} x^2-2^{\frac{n+3}{n}} \pi ^{1/n}\Big] \nonumber\\
f_6(x) &=& 32 \pi  B (3 a-5 b x)^{1/n} \Big[2 \pi  B n x (3 a-5 b x)^{1/n}- \nonumber\\
&& (8 \pi )^{1/n}\Big] .\nonumber
\end{eqnarray}
 The variation of interior metric function of the chosen compact star: X-ray pulsar LMC X-4 with distance $r$ is shown in Fig. \ref{fi1} and  the trends of the above physical quantities are depicted in Figs. \ref{fi2}-\ref{fi4}.

\begin{figure}[t]
\centering
\includegraphics[width=7cm,height=4.5cm]{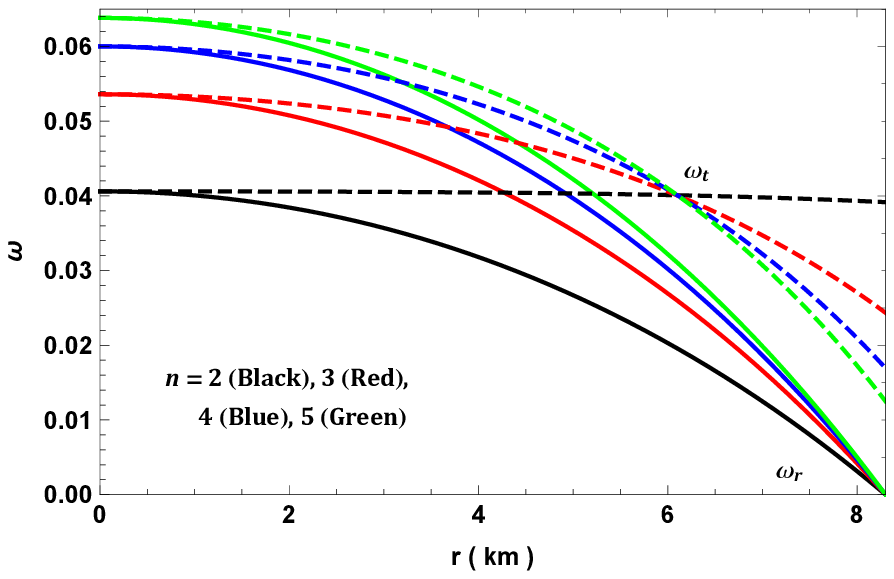}
\caption{Variation of equation of state parameters with $r$ for LMC X-4 assuming $M=1.04M_\odot,~R=8.301~km, ~a = 0.0039$ and $A = 0.7$.}\label{fi5}
\includegraphics[width=7cm,height=4.5cm]{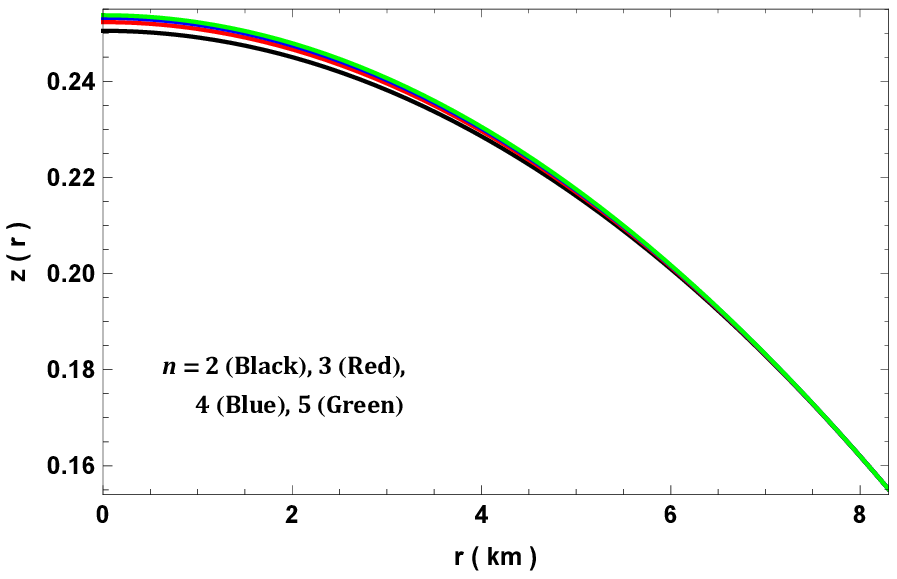}
\caption{Variation of redshift with $r$ for LMC X-4 assuming $M=1.04M_\odot,~R=8.301~km, ~a = 0.0039$ and $A = 0.7$.}\label{fi6}
\end{figure}

The mass, compactness parameter, equation of state parameter and red-shift can be determined as:
\begin{eqnarray}
m(r) &=& 4\pi \int r^2 \rho(r) ~dr=\frac{1}{2} \left(a r^3-b r^5\right)\\
u(r) &=& {2\mu(r) \over r}\\
\omega_r &=& {p_r \over \rho} ~~;~~\omega_t = {p_t \over \rho}\\
z(r) &=& e^{-\nu/2}-1={1 \over y(r)}-1.
\end{eqnarray}
It is to be noted that for a realistic physical system, the EoS parameters  must be less than unity, which is depicted in Fig. \ref{fi5}. The redshift profile is shown in Fig. \ref{fi6}.

\section{Non-singular nature of the solution}\label{4}
For a physical viable solution for an astrophysical compact object we must  ensure that the central values of density, pressure etc.are finite. {\it i.e}.
\begin{eqnarray}
\rho_c &=& {3 a \over 8\pi}  >0, \label{rhc}\\
p_{rc} &=& p_{tc} =  A \rho_c-B \rho_c^{1/n} > 0 \label{pc}
\end{eqnarray}
provided $a>0$ and $A>B\rho_c^{1/n-1}$. These inequalities provide a bound on the constant parameters and also implies that the solution is free from singularities.

\begin{figure}[t]
\centering
\resizebox{0.8\hsize}{!}{\includegraphics[width=7cm,height=4.5cm]{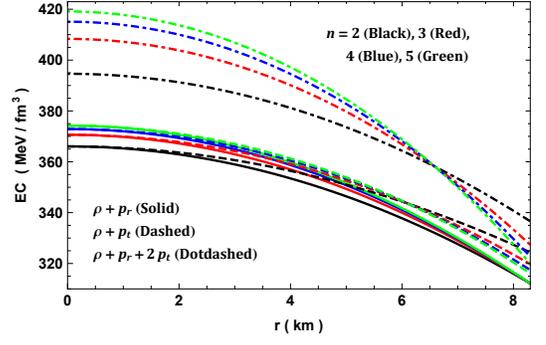}}
\caption{Variation of energy conditions with $r$ for LMC X-4 assuming $M=1.04M_\odot,~R=8.301~km, ~a = 0.0039$ and $A = 0.7$.}\label{fi7}
\end{figure}
\section{Boundary Conditions and determination of constants}\label{5}

Assuming the exterior space-time to be $Schwarzschild$'s which is given as
\begin{eqnarray}
ds^{2} &=& \left(1-\frac{2M}{r}\right)dt^{2}-\left(1-\frac{2M}{r}\right)^{-1}dr^{2} \nonumber \\
&& -r^{2}\big(d\theta^{2}+\sin^{2}\theta~ d\phi^{2} \big).
\end{eqnarray}

The continuity of the metric coefficients $e^\nu$ and $e^{-\lambda}$ across the boundary $r=R$, also imposing that the radial pressure vanishes at the boundary  yields:
\begin{eqnarray}
1-\frac{2M}{R} &=& e^{\nu_s} = e^{-\lambda_s}\label{b}\\
p_r(r=R) &=& 0 
\end{eqnarray}
that leads to
\begin{eqnarray}
b &=& \frac{a R^3-2 M}{R^5}\\\label{b1}
B &=& A (8 \pi )^{\frac{1}{n}-1} \left(3 a-5 b R^2\right)^{1-\frac{1}{n}}\\\label{b2}
F &=& \sqrt{1-\frac{2 M}{R}} \left(1-a R^2+b R^4\right)^{\frac{5 A}{8}+\frac{1}{8}} \nonumber \\
&& \exp \bigg[\frac{-a (A+1) }{4 \sqrt{4 b-a^2}} \tan ^{-1}\left(\frac{2 b R^2-a}{\sqrt{4 b-a^2}}\right)- \nonumber \\
&& \frac{2^{\frac{n-4}{n}} 5^{1/n} Bn \pi ^{\frac{n-1}{n}} \tau(R)  \left(3 a-5 b R^2\right)^{1/n}}{\sqrt{a^2-4 b}}\bigg]\,
\end{eqnarray}
where $M$ and $R$ denotes the mass and radius of the chosen compact star respectively, while $a$ and $A$ are kept as free parameter.

\section{Energy Conditions}\label{6}

Our next goal is to examine the condition under which static
stellar configurations, satisfies all the energy conditions  namely, weak energy condition (WEC), null energy condition (NEC), dominant energy condition (DEC) and  strong energy condition (SEC) at all points inside the star. The above energy conditions are determined by the following inequalities:

\begin{eqnarray}
\text{WEC} &:& T_{\mu \nu}t^\mu t^\nu \ge 0~\mbox{or}~\rho \geq  0,~\rho+p_i \ge 0 \\
\text{NEC} &:& T_{\mu \nu}l^\mu l^\nu \ge 0~\mbox{or}~ \rho+p_i \geq  0\\
\text{DEC} &:& T_{\mu \nu}t^\mu t^\nu \ge 0 ~\mbox{or}~ \rho \ge |p_i|\\
&& \mbox{where}~~T^{\mu \nu}t_\mu \in \mbox{nonspace-like vector} \nonumber \\
\text{SEC} &:& T_{\mu \nu}t^\mu t^\nu - {1 \over 2} T^\lambda_\lambda t^\sigma t_\sigma \ge 0 ~\mbox{or} \nonumber \\
&& \rho+\sum_i p_i \ge 0 ~~~\rho+p_i \ge 0
\end{eqnarray}
where $i\equiv (radial~r, transverse ~t),~t^\mu$ and $l^\mu$ are time-like vector and null vector respectively. The verification of the energy conditions are shown in Fig. \ref{fi7}.
\begin{figure}[t]
\centering
\resizebox{0.8\hsize}{!}{\includegraphics[width=7cm,height=4.5cm]{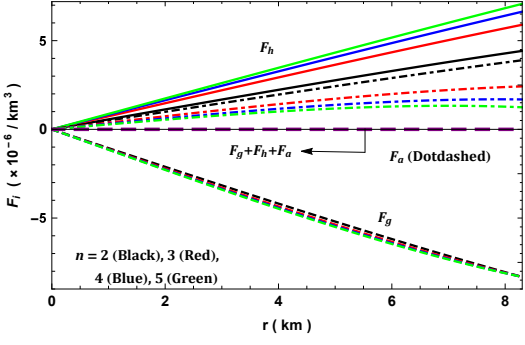}}
\caption{Variation of forces in TOV-equation with $r$ for LMC X-4 assuming $M=1.04M_\odot,~R=8.301~km, ~a = 0.0039$ and $A = 0.7$.}\label{fi8}
\end{figure}

\section{Stability of the model and equilibrium}\label{7}

\subsection{Equilibrium under various forces}

The generalized Tolman-Oppenheimer-Volkoff (TOV)  equation determine whether a relativistic stellar system is in equilibrium or not. Mathematically, it is given by
\begin{equation}
-\frac{M_g(r)(\rho+p_r)}{r}e^{(\nu-\lambda)/2}-\frac{dp_r}{dr}+\frac{2}{r}(p_t-p_r)=0, \label{to1}
\end{equation}
where $M_g(r) $ is the gravitational mass and is calculated using the Tolman-Whittaker formula and the Einstein field equations. The expression is given as
\begin{eqnarray}
M_g(r) &=& 4 \pi \int_0^r \big(T^t_t-T^r_r-T^\theta_\theta-T^\phi_\phi \big) r^2 e^{(\nu+\lambda)/ 2}dr. \nonumber \\ \label{mg}
\end{eqnarray}
For the Eqs. (\ref{dens})-(\ref{prt}), the above Eqn. (\ref{mg}) reduces to
\begin{equation}
M_g(r)=\frac{1}{2}re^{(\lambda-\nu)/2}~\nu'.
\end{equation}
Using the above expression of $M_g(r)$ in Eqn.(\ref{to1}), we get
\begin{equation}\label{tov}
-\frac{\nu'}{2}(\rho+p_r)-\frac{dp_r}{dr}+\frac{2}{r}(p_t-p_r)=0
\end{equation}
which can also be expressed as:
\begin{equation}
F_g+F_h+F_a=0,
\end{equation}
where $F_g,~ F_h$ and $F_a$ are the gravitational, hydrostatics and anisotropic forces respectively. {\it i.e.}
\begin{eqnarray}
F_g &=& -\frac{\nu'}{2}(\rho+p_r)~,~~F_h = -\frac{dp_r}{dr} ~,~~
F_a = {2\Delta \over r}.\label{for}
\end{eqnarray}

Variation of the above forces with distance $r$ is shown in Fig. \ref{fi8} which clearly convinces that the solution is in equilibrium.

\begin{figure}[t]
\centering
\resizebox{0.8\hsize}{!}{\includegraphics[width=7cm,height=4.5cm]{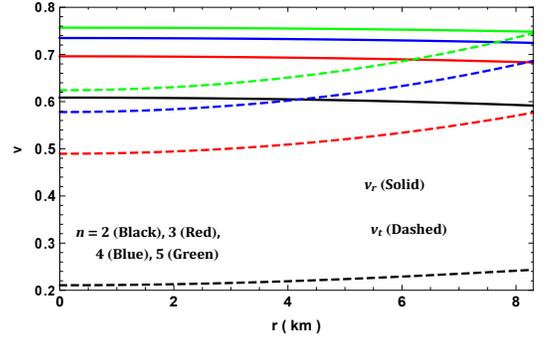}}
\caption{Variation of speed of sound with $r$ for LMC X-4 assuming $M=1.04M_\odot,~R=8.301~km, ~a = 0.0039$ and $A = 0.7$.}\label{fi9}
\end{figure}

\subsection{Causality and stability condition}
\begin{figure}[t]
\centering
\resizebox{0.8\hsize}{!}{\includegraphics[width=7cm,height=4.5cm]{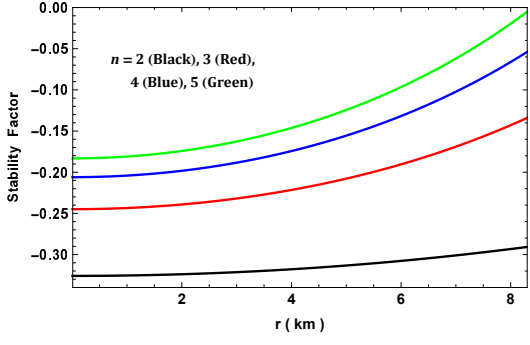}}
\caption{Variation of stability factor with $r$ for LMC X-4 assuming $M=1.04M_\odot,~R=8.301~km, ~a = 0.0039$ and $A = 0.7$.}\label{fi10}
\end{figure}

The sound speed is an important parameter to check the causality condition, which  implies that the radial velocity $(v_{r}^{2})$ and tangential velocity $(v_{t}^{2})$ of sound should be less than unity everywhere within the compact object, {\it i.e.}, $ 0 < v^2_r\leq 1$ and $0 < v^2_t \leq 1$, where
\begin{eqnarray}
v_{r}^{2} = {dp_r \over d\rho},~~v_{t}^{2} = {dp_t \over d\rho}.
\end{eqnarray}
Figure \ref{fi9} verify the subliminal sound speed at the interior. The solution also satisfy the Abreu's stability criterion \cite{Abreu}, which states: $-1 \le v_t^2-v_r^2 \le 0$, which is depicted in Fig. \ref{fi10}.

\subsection{Adiabatic index and stability condition}
For an anisotropic matter distribution adiabatic index also determine the stability of the fluid distribution which is defined as \cite{cha93},
\begin{equation}
\Gamma_r=\frac{\rho+p_r}{p_r}\frac{dp_r}{d\rho}.
\end{equation}

\begin{figure}[t]
\centering
\resizebox{0.8\hsize}{!}{\includegraphics[width=7cm,height=4.5cm]{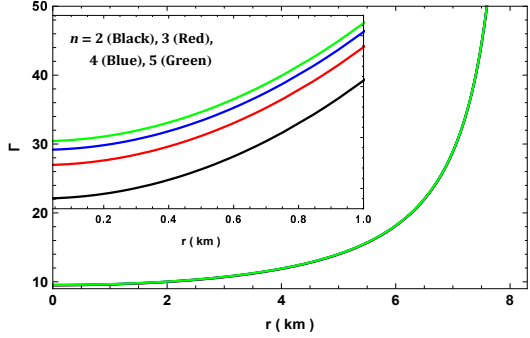}}
\caption{Variation of adiabatic index with $r$ for LMC X-4 assuming $M=1.04M_\odot,~R=8.301~km, ~a = 0.0039$ and $A = 0.7$.}\label{fi11}
\end{figure}

According to Bondi condition \cite{Bondi} $\Gamma_r>4/3$ gives a stable Newtonian system whereas $\Gamma =4/3$ is the condition for a neutral equilibrium. This condition is partially valid for anisotropic case as it depends on nature of anisotropy. 
For an anisotropic fluid sphere the adiabatic index modify to \cite{cha93},
\begin{equation}
\Gamma>\frac{4}{3}+\left[\frac{4}{3}\frac{(p_{ti}-p_{ri})}{|p_{ri}^\prime|r}+\frac{8\pi}{3}\frac{\rho_ip_{ri}}{|p_{ri}^\prime|}r\right]_{max},
\end{equation}
where, $p_{ri}$, $p_{ti}$, and $\rho_i$ are the initial radial, tangential pressures and energy density respectively in static equilibrium. Within the square bracket, first term gives the anisotropic modification and the last term is relativistic correction to $\Gamma$ \cite{cha93, her92}. If the anisotropy is positive, then a system with $\Gamma_r  > 4/3$ may not be in stable and vice versa. For such case the adiabatic index is more than 4/3 including the extra correcting terms (see Fig. \ref{fi11}) and therefore is stable.


\begin{figure}[t]
\centering
\resizebox{0.8\hsize}{!}{\includegraphics[width=7cm,height=4.5cm]{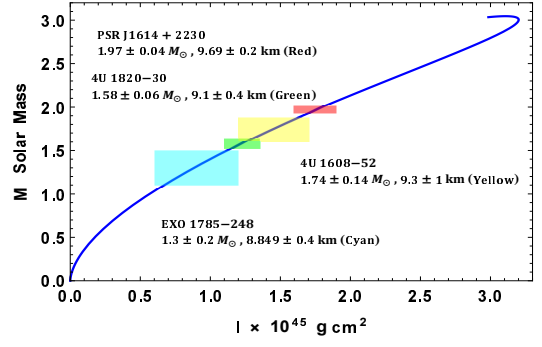}}
\caption{Variation of moment of inertia with $M$ ($A=0.72, ~B=0.01025,~n=2,~b=8\times 10^{-7}$).}\label{fi13}
\end{figure}

\subsection{Harrison-Zeldovich-Novikov static stability criterion}\label{10}

\begin{figure}[t]
\centering
\resizebox{0.8\hsize}{!}{\includegraphics[width=7cm,height=4.5cm]{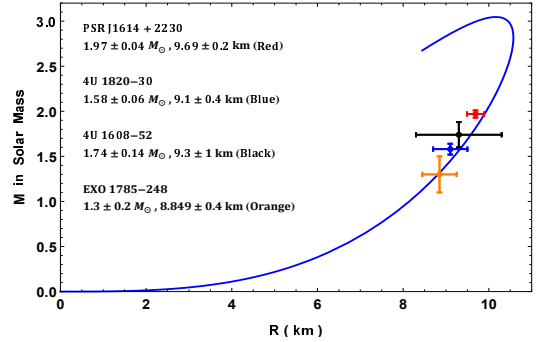}}
\caption{$M-R$ curve fitted for few compact stars ($A=0.72, ~B=0.01025,~n=2,~b=8\times 10^{-7}$).}\label{fi12}
\end{figure}

The static stability criterion proposed by Harrison {\it et al}., \cite{har65} and Zeldovich-Novikov \cite{zel71} are much simpler than the Chandrasekhar's criterion \cite{chan64}.
It states that any stellar model is a stable configuration only if its mass increases with growing central density i.e. $d\mu /d\rho_c > 0$. The opposite inequality {\it i.e.,} $d\mu /d\rho_c < 0$ always implies instability of stellar models. Mass as a function of central density is given below:
\begin{eqnarray}
\mu (\rho_c) &=& \frac{1}{6} \left(8 \pi  R^3 \rho_c-3 b R^5\right)\label{mrhc}
\end{eqnarray}
One can clearly see the mass is a linear function of its central density which straight ways yields $\partial \mu(\rho_c)/\partial \rho_c>0$ at constant $R$  since $b$ is of the order of $10^{-5}$. Therefore, the static stability criterion is fulfilled.

\subsection{Determination of moment of inertia under slow rotation approximation}

 Bejger and Haensel \cite{bej} defined an approximate expression for the moment of inertia of compact stars under slow rotation which is within 5\% accuracy given by
\begin{equation}
I = {2 \over 5}\left(1+{M \over R} \cdot {km \over M_\odot}\right)MR^2.
\end{equation}
The $M-I$ curve in Fig. \ref{fi13} can be generated using the above expression incorporating the variations of mass and radius from $M-R$ curve in Fig.\ref{fi12}. As per Fig. \ref{fi13}, the estimated values of $I$ are $1.75\times 10^{45} g~cm^2$ (PSR J1614 + 2230), $1.42\times 10^{45} g~cm^2$ (4U 1608-52), $1.23\times 10^{45} g~cm^2$ (4U 1820-30) and $8.92\times 10^{44} g~cm^2$ (EXO 1785-248).

\section{Discussion and conclusion}\label{dis}
We have studied the effects of exotic matter on the astrophysical compact objects. Owing to the high density in the interior of the compact objects, it is quite possible that many exotic form of matter may exist. To explore this idea we employ Tolman VII solutions to  a generalized non-linear  EoS. of the form mentioned in Eqn.(\ref{zeo}). A special form of such EoS has been studied widely in case ever existing universe \cite{Dadhich}, where the values of parameter determine the primordial constitutes of universe. We have modeled such constituents in a stellar compact star configuration, namely an X-ray pulsar i.e., LMC X-4, having the observed mass as $M = 1.04 M\odot$. In order to solve Einstein's equation and derive different thermodynamic properties from it we first need to constraint the parameters of the model. The constraint equations of the parameters are deduced from the boundary conditions which are shown in Eqns.(\ref{b1}) and (\ref{b2}). Here parameters $b$ and $B$ are related to parameters $n$, $A$ and $a$, which are treated as free parameters. We explore several physical aspects of the model analytically along with graphical display in order to verify that the model can depict a viable astrophysical compact object. Our analysis show that model is free from all singularities and is stable for the parameter $A$ ranging from $0.58-1$ for all $n$. Therefore, we choose the value of $A$ in this range and investigate  all the physical aspects of the model for different values of $n$, which are discussed below:

The metric potentials as a function of $r$ is displayed in Fig. \ref{fi1}. We notice that both the potentials are finite at the stellar center. The metric potential $e^\nu$ at center($r=0$) is constant and is monotonically increases towards the boundary  whereas $e^{-\lambda}$ at $r=0$ is one and is monotonically decreasing towards the boundary of the star. They are free from singularities inside the star and  values of both the potential is identical at the surface of the star.

\begin{figure}[t]
\centering
\resizebox{0.9\hsize}{!}{\includegraphics[width=7cm,height=4.5cm]{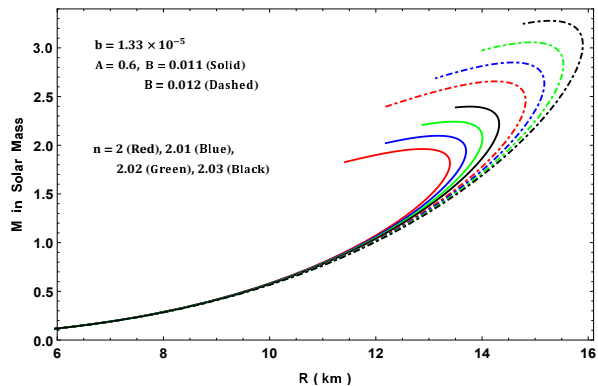}}
\caption{Effects on $M-R$ curve due to $B$ and $n$ parameter.}\label{fi14}
\end{figure}

The thermodynamic properties of the interior of the star is shown in Fig. \ref{fi2} and Fig. \ref{fi3}. From the plots we notice that the density $\rho$ and pressures $p_r$ and $p_t$ are positive and maximum at the core of the star. They decreases towards the surface of the star and are free from any central singularity. The radial pressure $p_r$ drops to zero at the boundary of the star but the tangential pressure $p_t$ remains finite both at the core and at the boundary. However the values of both $p_r$ and $p_t$ are equal at the center of the star which implies that the pressure anisotropy factor vanishes at the center, $\Delta(r = 0) = 0$ as is evident from the profile of anisotropy $\Delta$ in Fig. \ref{fi4}. Also one can note from it that anisotropy increases with distance $r$ and is maximum at the surface of the star. Moreover the anisotropy decreases for increasing values of $n$.  $\Delta >0$ inside the star causes the anisotropic force to be repulsive in nature. One important information extracted from these graphs is that as the parameter $n$ increases the central values of pressure increases and anisotropy reduces. This means that the EoS is getting stiffer as $n$ increase i.e.  by increasing dark energy/exotic contribution ($B\rho^{1/n}$) makes the EoS stiffer.

This can also be confirms from the variation of both radial $(\omega_r)$ and transverse $(\omega_t)$ EoS parameters with radial distance $r$ (Fig. \ref{fi5}). As $n$ increases the $\omega-$parameters increases. The variation of gravitational redshift with radial distance is shown in Fig.(\ref{fi6}), which shows it is monotone decreasing. All the energy conditions mentioned in Section (\ref{6}) are fulfilled by the stellar configuration which is depicted in Fig. \ref{fi7}. This suggests that the model is viable in nature with no instability or presence of ghost (negative mass or energy). 

Equilibrium of a stellar system is determined by different forces that generate inside the system. These forces can be estimated from the TOV Eqns.(\ref{tov})-(\ref{for}).The variation of these forces, namely gravitational force $(F_g)$, hydrodynamic force $(F_h)$, and anisotropic force $(F_a)$ are shown in Fig.(\ref{fi8}). From the figure we conclude that the system is in equilibrium. Another important criterion to test the stability is the causality condition which is valid when the sound is less than one. Fig. \ref{fi9} confirms not only the fulfillment of the causality condition but also the increase in stiffness as $n$ increases. It also follows the Abereu's stability criterion {\it viz}: $-1 \le v_t^2-v_r^2 \le 0$ as can be seen from Fig. \ref{fi10}.

Stability of anisotropic matter distribution is also determined by the adiabatic index of the constituents of the stellar configuration, which is verified in Fig. \ref{fi11}. For the values of $n$ considered the anisotropic stellar configuration is steady as $\Gamma > 4/3$ in the interior of the system. Lastly the Harrison-Zeldovich-Novikov static stability criterion is also checked for the model. Figure depicts the fulfillment of the criteria as we see the mass increases with growing central density which is the necessary condition for a stable stellar configuration.

To verify whether the stellar configuration can depict more varieties of observed compact objects we construct the $M-R$ plot  in Fig. \ref{fi12}. From this curve we notice that many observed compact objects can be modeled considering exotic matter as their constituents. Particularly, it is well fitted within the observation errors for EXO 1785-248 \cite{gang}, PSR J1614 + 2230 \cite{demo}, 4U 1608-52 \cite{gu} and 4U 1820-30 \cite{guv}. The approximate moment of inertia $I$ for these fitted compact objects can be estimated  form Fig. \ref{fi13} within $5\%$ accuracy. The $M-R$ curve in Fig. \ref{fi14} demonstrates the effect of $B$ and $n$ parameters. As $n$ increase the stiffness increases resulting into increase in $M_{max}$ significantly. In the similar fashion, a small change in $B$ increases $M_{max}$ significantly. It means as $n$ and $B$ increases the dark energy contribution increases resulting into more stiffer EoS. We therefore conclude that the theoretical prediction of stellar configuration with non-linear EoS which incorporates the dark energy/exotic matter  follows all the stability criterion as well as goes well with the observational data.

\section*{Acknowledgments}
 AA is thankful to UGC for providing financial support under the scheme Dr. D.S. Kothari postdoctoral fellowship. FR is thankful to the authority of Inter-University Centre for Astronomy and Astrophysics, Pune, India for providing Associateship. We are grateful to the referees for their valuable comments and fruitful suggestions.


\end{document}